 \documentclass[aps,prl,twocolumn,groupedaddress, amsmath, amssymb, showpacs]{revtex4-1}
 
\newcommand{\B}{\textrm{B}}

\newcommand{\vc}{\mathbf}
\usepackage{graphicx}
\usepackage{color}

\begin{document}


\title{An Effective Potential Theory for Transport Coefficients Across Coupling Regimes }


\author{Scott D.\ Baalrud$^{1,2}$}
\author{J\'{e}r\^{o}me Daligault$^1$}
\affiliation{$^1$Theoretical Division, Los Alamos National Laboratory, Los Alamos, New Mexico 87545, USA}
\affiliation{$^2$Department of Physics and Astronomy, University of Iowa, Iowa City, Iowa 52242, USA}


\date{\today}

\begin{abstract}

A plasma transport theory that spans weak to strong coupling is developed from a binary collision picture, but where the interaction potential is taken to be an effective potential that includes correlation effects and screening self-consistently. This physically motivated approach provides a practical model for evaluating transport coefficients across coupling regimes. The theory is shown to compare well with classical molecular dynamics simulations of temperature relaxation in electron-ion plasmas, as well as simulations and experiments of self-diffusion in one component plasmas. The approach is versatile and can be applied to other transport coefficients as well. 

\end{abstract}




\maketitle


The microscopic dynamics of Coulomb collisions determines macroscopic transport properties of plasmas such as diffusivity, resistivity, viscosity, etc.~\cite{chap:70,brag:65}.  Usually plasmas are so hot and dilute that the average particle kinetic energy greatly exceeds the potential energy of interaction. In this weakly coupled regime, Coulomb collisions consist of a series of many small angle binary scattering events~\cite{land:36,spit:53,rose:57,lena:60}. Strongly coupled plasmas are fundamentally different. In this regime, the interaction potential energy exceeds the particle kinetic energies, so scattering angles are large and correlation effects are important.  Plasmas in several modern experiments, including inertial confinement fusion (ICF) \cite{atze:04,lee:84}, antimatter plasmas \cite{andr:10}, ultra cold plasmas \cite{kill:07} and dusty plasmas \cite{merl:04}, exhibit strong coupling effects; as do some naturally occurring objects including neutron star crusts~\cite{pote:99}, white dwarf stars~\cite{much:84,iben:85} and giant planet interiors~\cite{lore:09,chab:06,dense}.  

Understanding how transport properties are modified in strongly coupled plasmas is interesting both from a fundamental physics standpoint and as a practical matter. Accounting for correlation effects remains a challenge for theory, even though accurate transport coefficients are critical input to the macroscopic (fluid) equations used to model these systems.  Transport calculations typically rely on computationally expensive particle simulations, such as molecular dynamics (MD)~\cite{baus:80,desj:02,donk:09}. Analytic theory is desirable because it can both elucidate the physical processes that influence transport at strong coupling, and provide an efficient means for estimating the transport coefficients that fluid equations require as input~\cite{rose:95,rein:95,bene:12}.  In this Letter, we describe a physically motivated method of extending conventional transport calculations, which is efficient enough to be practically implemented in fluid simulation codes. The theory provides coefficients that agree with experimental \cite{bann:12} and classical MD simulation data \cite{dimo:08,dali:12a} across coupling regimes.

Like weakly coupled theories, our theory is based on a binary collision picture, but where particles interact via an effective potential that includes average effects of the intervening medium; including both correlations and screening. This effective potential is used to derive a scattering cross section, which is then applied to the Boltzmann collision operator and Chapman-Enskog collision integrals~\cite{chap:70} to calculate the various transport coefficients. 

In fact, traditional plasma theories also rely on an effective potential. The bare Coulomb potential neglects screening of the intervening medium and leads to a divergent collision operator. To fix this unphysical divergence, Landau utilized the weak coupling assumption to cutoff the impact parameter at the Debye screening length~\cite{land:36}. Hence, imposing an effective potential. This approximation leads to the traditional Coulomb logarithm, $\ln \Lambda$, where $\Lambda \sim \Gamma^{-3/2}$ is the plasma parameter. It is valid in the limit that this parameter is asymptotically large. Here, the Coulomb coupling parameter, $\Gamma = Z^2 e^2/(k_\B T a)$ where $a = (4\pi n/3)^{-1/3}$ is the Wigner-Seitz radius, will be used to quantify coupling strength. The Lenard-Balescu equation is an alternative plasma kinetic theory derived from the BBGKY hierarchy~\cite{lena:60}. It has the advantage of accounting for screening self-consistently, but it does not account for close interactions and also diverges. Again, this divergence is fixed through the weak coupling approximation. Using the screened Coulomb (Yukawa) potential as an effective potential avoids the logarithmically divergent integrals and can extend the binary collision approach to larger coupling strength~\cite{libo:59,maso:67,paqu:86,baal:12,geri:02}, but it does not capture correlation effects, which onset when $\Gamma \gtrsim 1$. Can the binary collision picture be extended further by using an effective interaction potential that accounts for correlation effects in addition to screening? In this Letter, we present evidence that it can. 

Previous theories of transport in strongly coupled plasmas have largely focused on developing new closure schemes of the BBGKY hierarchy that include correlations~\cite{ichi:92,hans:06,dali:09,dali:11,boer:82}. These typically either derive a new collision operator that has a generalized linear dielectric response with local field corrections \cite{ichi:92,dali:09}, or calculate transport properties from higher-order equilibrium correlation functions~\cite{hans:06,boer:82}. The salient feature of these closures is that they go beyond the mean field approximation of conventional plasma theories to include correlation effects. The mean field approximation is justified in the weakly coupled limit because there are many particles within the interaction length scale (Debye length) of a test particle.  However, in a strongly coupled plasma the interaction distance is instead characterized by the inter-particle spacing. In this regime, the test particle self interaction must be neglected, and correlations accounted for.   

Next, we establish a relationship between the effective interaction potential ($\phi$) and pair correlation function ($g_2$). This enables determination of $\phi$ from closures that include correlation effects in $g_2$.  To illustrate this point, consider the second BBGKY equation for $g_2$ for a classical one component system
\begin{subequations}
\begin{align}
&\frac{\partial g_2(1,2)}{\partial t} = \left[L_1^0+L_2^0\right]g_2(1,2)+L_{12}f(1)f(2)   \label{1}\\ 
&+L_{12}g_2(1,2)   \label{l2}\\
&+\int{d3\left[ L_{13}f(1)g_2(2,3)+L_{13}f(3)g_2(1,2) + (1\leftrightarrow 2) \right]}   \label{l3}\\ 
&+\int{d3 (L_{13}+L_{23})g_3(1,2,3)}   \label{l4} 
\end{align}
\end{subequations}
where $L_i^0 = - \vc{v}_i \cdot \nabla_i$, $L_{i,j} = \nabla v_{i,j} \cdot \partial_{i,j}$ and $v_{i,j} = v( |\vc{r}_i - \vc{r}_j |)$ is the bare Coulomb potential. The usual kinetic theories can be obtained by neglecting certain terms in Eq.~(1). The Landau collision operator is obtained by neglecting terms (\ref{l2})--(\ref{l4}), the Lenard-Balescu collision operator by neglecting (\ref{l2}) and (\ref{l3}), and the Boltzmann collision operator by neglecting (\ref{l3}) and (\ref{l4}). Each choice defines an approximation for $g_2$ and, in turn, a different collision operator.

For our purposes, the equilibrium limit of these approximations is instructive. At equilibrium, $g_2 (1,2) = n^2 f_\textrm{M} (\vc{p}_1) f_\textrm{M} (\vc{p}_2) h(| \vc{r}_1 - \vc{r}_2|)$ where $f_\textrm{M}$ is a Maxwellian, $h(r) = g(r) -1$ and $g(r)$ is the pair distribution function; $ng(r)$ is the average density at a distance $|\vc{r}|$ from any particle. In this limit, the Landau and Lenard-Balescu closures correspond to assuming $g^\textrm{L}(r) = 1 - ev(r)/k_\B T$ and $g^{\textrm{LB}}(r) = 1 - e \phi_\textrm{sc}(r)/k_\B T$ where $\phi_{sc} = q \exp(-r/\lambda_D)/r$ is the screened Coulomb potential. These can be obtained from the weakly coupled limit ($e\phi/k_\B T \ll 1$) of the general equilibrium relationship~\cite{hill:60}
\begin{equation}
g(r) = \exp (-e\phi/k_\B T) , \label{eq:grphi}
\end{equation}
where 
\begin{equation}
- \nabla \phi =   \frac{\int e^{-U/k_\B T} (-\nabla_1 U)d \vc{r}_3 \ldots d \vc{r}_N}{\int e^{-U/k_\B T} d \vc{r}_3 \ldots d \vc{r}_N}  \label{eq:epdef}
\end{equation}
defines an effective interaction potential~\cite{epnote}, and $U=\sum_{i,j} v_{i,j}$ is the total interaction energy. The quantity $- \nabla \phi$ represents the mean force acting on particle 1, with particles 1 and 2 held at fixed positions ($\vc{r}_1$ and $\vc{r}_2$), averaged over the positions of all other particles. The usual plasma theories rely on the $e\phi/k_\B T \ll 1$ assumption, whereas the Boltzmann collision operator does not. Actually, the Boltzmann collision operator corresponds to using the bare Coulomb potential ($\phi = v$) in Eq.~(\ref{eq:grphi}), which neglects both screening and correlations. We will use an effective potential that includes these effects. 

\begin{figure}
\includegraphics[width=8.0cm]{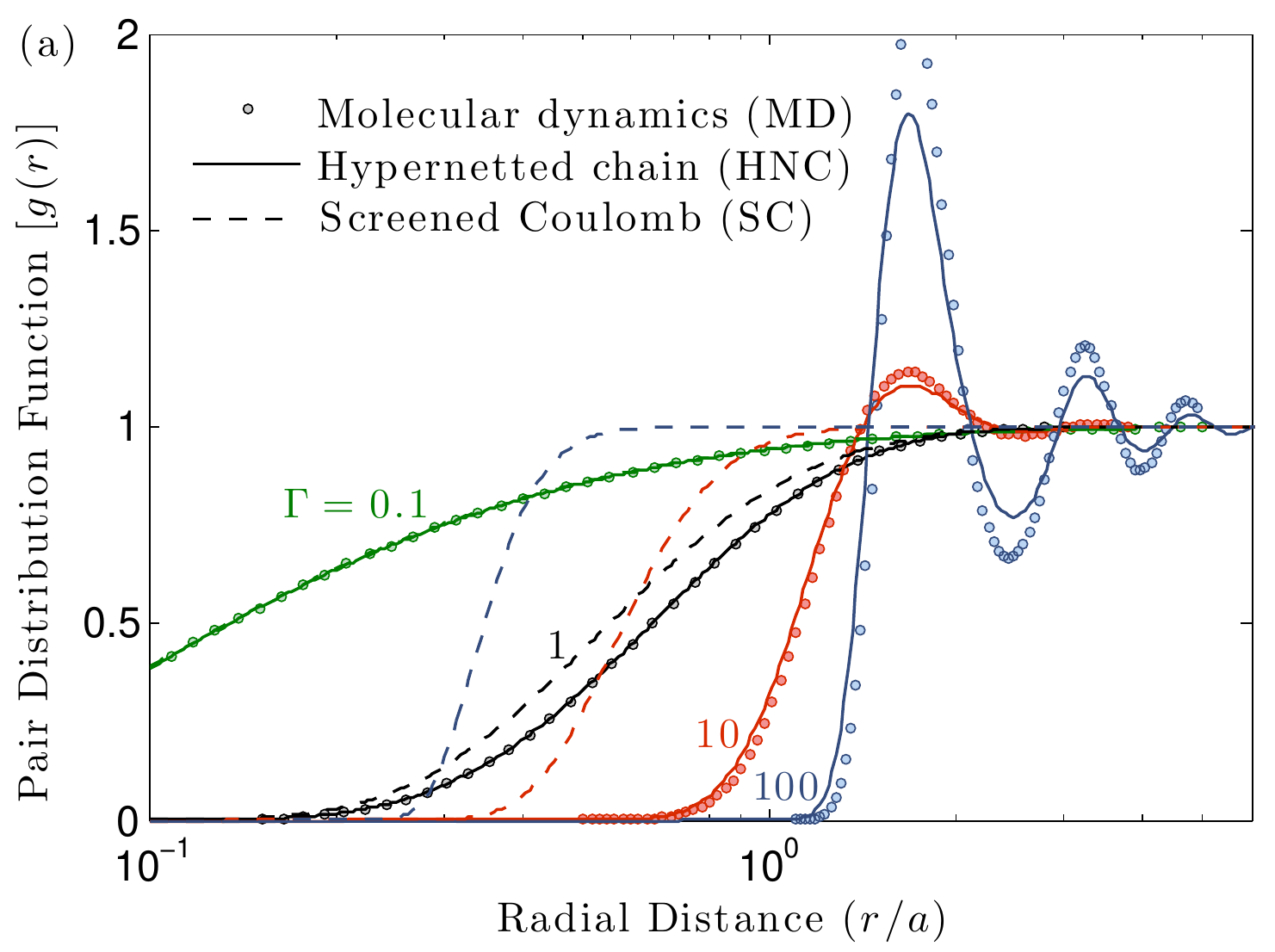}
\includegraphics[width=8.0cm]{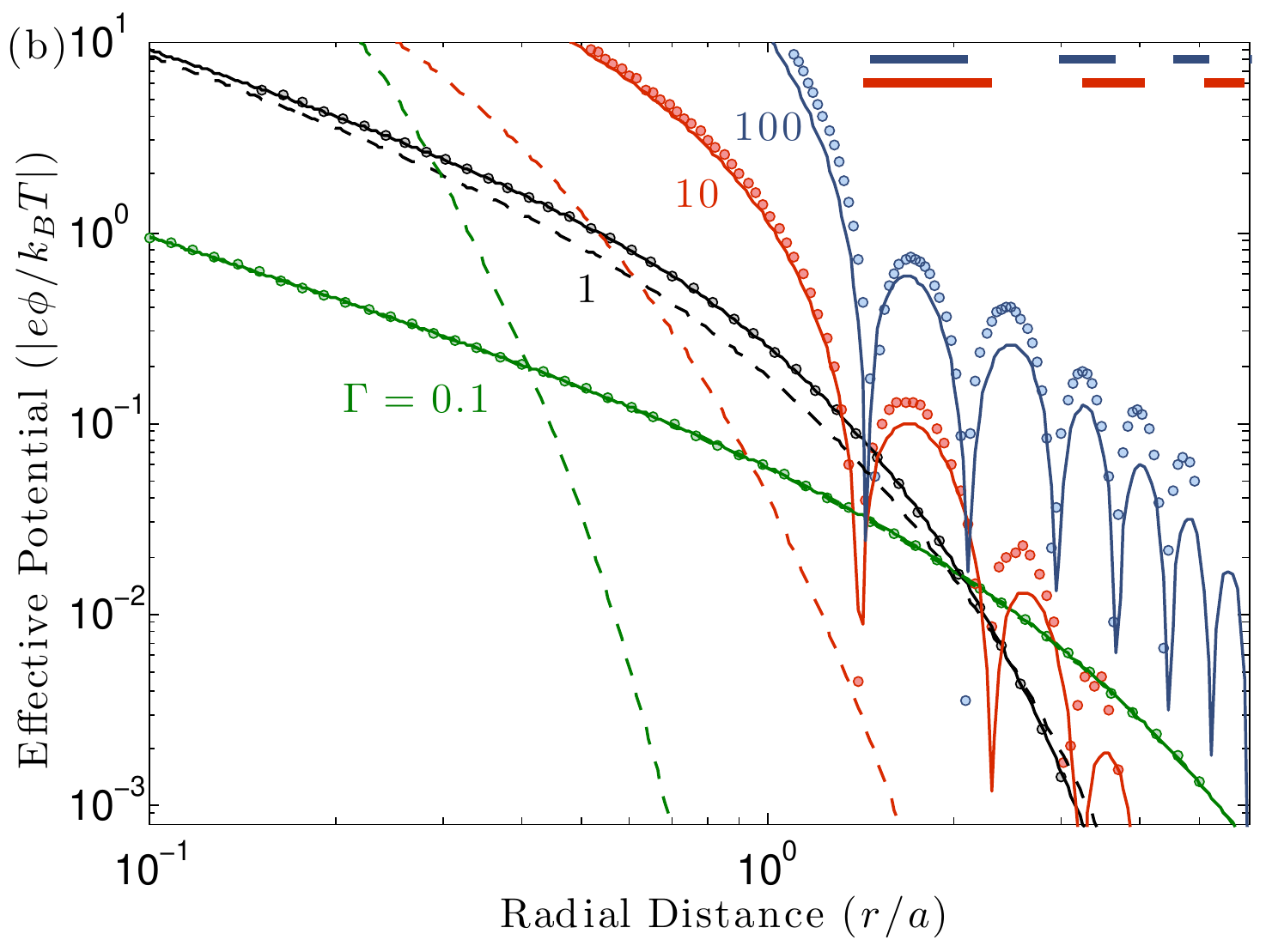}
\caption{(a) Pair distribution function for the OCP determined from classical MD (circles), HNC (solid lines) and the screened Coulomb potential (dashed lines) for $\Gamma = 0.1, 1, 10$ and $100$.  (b) Magnitude of the effective interaction potential calculated from Eq.~(\ref{eq:grphi}). The thick red and blue lines show the $r/a$ intervals where $\phi < 0$ for $\Gamma =10$ and $100$. }
\label{fg:gr}
\end{figure}

Figure~\ref{fg:gr}a shows a comparison of $g(r)$ obtained from different approximations with that extracted from classical MD simulations of a one component plasma (OCP)~\cite{dali:12a}. Figure~\ref{fg:gr}b shows the corresponding effective potential from Eq.~(\ref{eq:grphi}). The screened Coulomb potential is an excellent approximation in the weakly coupled regime, $\Gamma \ll 1$, but this breaks down as correlation effects onset at $\Gamma \gtrsim 1$. It fails entirely in the strongly correlated cases. To obtain an analytic approximation for $g(r)$ that includes correlation effects, but doesn't rely on computationally expensive MD simulations, we use the hypernetted chain (HNC) closure. HNC is a well-established approximation in which $g(r)$ is determined from the two coupled equations \cite{hans:06}
\begin{equation}
g(\vc{r}) = \exp \biggl[ - v(\vc{r})/k_\B T + n \int c(|\vc{r} - \vc{r}^\prime |) h(\vc{r}^\prime) d \vc{r}^\prime \biggr]  \label{eq:hncgr}
\end{equation}
and 
\begin{equation}
\hat{h}(\vc{k}) = \hat{c}(\vc{k})[ 1 + n \hat{h}(\vc{k})]  .  \label{eq:hnchk}
\end{equation}
For the OCP, $v(\vc{r})/k_\B T =  \Gamma a /r$ and $\hat{h} (\vc{k})$ denotes the Fourier transform of $h(\vc{r})$. Figure~\ref{fg:gr} shows that, like the screened Coulomb potential, HNC provides an excellent approximation for weak coupling. However, unlike the screened Coulomb, this agreement extends to the strongly coupled regime. Next, we apply this effective potential to determine transport properties. 

In the Chapman-Enskog theory, transport coefficients arise through the collision integrals \cite{chap:70}
\begin{equation}
\Omega_{ss^\prime}^{(l,k)} = \sqrt{\pi} \bar{v}_{ss^\prime} \int_0^\infty d\xi \xi^{2k+3} e^{-\xi^2} \int_0^\pi  d\theta \sigma_{ss^\prime} \sin \theta  (1 - \cos^l \theta) 
\end{equation}
in which $\theta$ is the scattering angle, $\xi = u/\bar{v}_{ss^\prime}$, $\bar{v}_{ss^\prime}^2 = v_{Ts}^2 + v_{Ts^\prime}^2$, $v_{Ts}^2 = 2T_s/m_s$, $u = | \vc{v} - \vc{v}^\prime |$ and $\sigma_{ss^\prime}$ is the differential scattering cross section. Here, $s$ and $s^\prime$ denote species. Alternatively, these can be written~\cite{baal:12}
\begin{equation}
\Omega_{ss^\prime}^{(l,k)} = \frac{3}{16} \frac{m_s}{m_{ss^\prime}} \frac{\nu_{ss^\prime}}{n_{s^\prime}} \frac{\Xi_{ss^\prime}^{(l,k)}}{\Xi_{ss^\prime}} , \label{eq:cints}
\end{equation}
where 
\begin{equation}
\Xi_{ss^\prime}^{(l,k)} = \frac{1}{2} \int_0^\infty d\xi\, \xi^{2k+3} e^{-\xi^2} \bar{\sigma}_{ss^\prime}^{(l)} / \sigma_o \label{eq:xilk}
\end{equation}
is a ``generalized Coulomb logarithm'' associated with the $(l,k)^\textrm{th}$ collision integral. Here, $\Xi_{ss^\prime} = \Xi_{ss^\prime}^{(1,1)}$ is the lowest order term, 
\begin{equation}
\nu_{ss^\prime} \equiv \frac{16 \sqrt{\pi} q_s^2 q_{s^\prime}^2 n_{s^\prime}}{3 m_s m_{ss^\prime} \bar{v}_{ss^\prime}^3} \Xi_{ss^\prime} \label{eq:nu}
\end{equation}
is a reference collision frequency,
\begin{equation}
\bar{\sigma}_{ss^\prime}^{(l)} = 2 \pi \int_0^\infty db\, b [1 - \cos^{l} (\pi - 2 \Theta) ]  \label{eq:sigl}
\end{equation}
is the $l^{\textrm{th}}$ momentum-transfer cross section, $\sigma_o = (\pi q_s^2 q_{s^\prime}^2)/(m_{ss^\prime}^2 \bar{v}_{ss^\prime}^4)$ is a reference cross section, and $m_{ss^\prime} =m_s m_{s^\prime}/(m_s + m_{s^\prime})$ is the reduced mass. The scattering angle is that of a classical binary collision 
\begin{equation}
\Theta = b \int_{r_o}^\infty dr\, r^{-2} [1 - b^2/r^2 - 2e\phi (r)/(m_{ss^\prime} u^2)]^{-1/2}  \label{eq:theta}
\end{equation}
in which $r_o$ is the distance of closest approach, determined from the largest root of the denominator in Eq.~(\ref{eq:theta}). 

Equations~(\ref{eq:cints})-(\ref{eq:theta}) determine the transport coefficients. First, we recover the weakly coupled limit. Applying the screened Coulomb potential provides generalized Coulomb logarithms that avoid the traditional divergences~\cite{libo:59,maso:67,paqu:86,baal:12}. For weak to moderate coupling, the lowest order Coulomb logarithm is $\Xi^{(1,1)} = \exp(\Lambda^{-1}) E_1 (\Lambda^{-1})$, where $E_1$ is the exponential integral~\cite{baal:12}. The weak coupling limit of this returns the conventional Coulomb logarithm including an order unity correction 
\begin{equation}
\Xi_{ss^\prime}^{(1,1)} \rightarrow \ln \Lambda -\gamma = \ln (0.56 \Lambda ) . \label{eq:xiwc}
\end{equation}
This order unity correction extends the conventional $\ln \Lambda$ solution to the moderate coupling regime $\ln \Lambda \gtrsim 2$. It has also been obtained by others using complicated renormalization techniques~\cite{aono:68,lifs:81,brow:05}. The result that is usually cited is $\ln (0.765 \Lambda)$. Equation~(\ref{eq:xiwc}) reduces to $\ln (0.79 \Lambda)$ in the limit $T_e=T_i$ and $\Lambda$ is defined using the electron Debye length, which is within $\sim 3 \%$ of the coefficient from previous calculations. Next, we extend these calculations into the strong coupling regime using the HNC obtained effective potential to compute the self diffusion coefficient for a OCP and the temperature relaxation rate of an electron-ion plasma. 

\begin{figure}
\includegraphics[width=8.0cm]{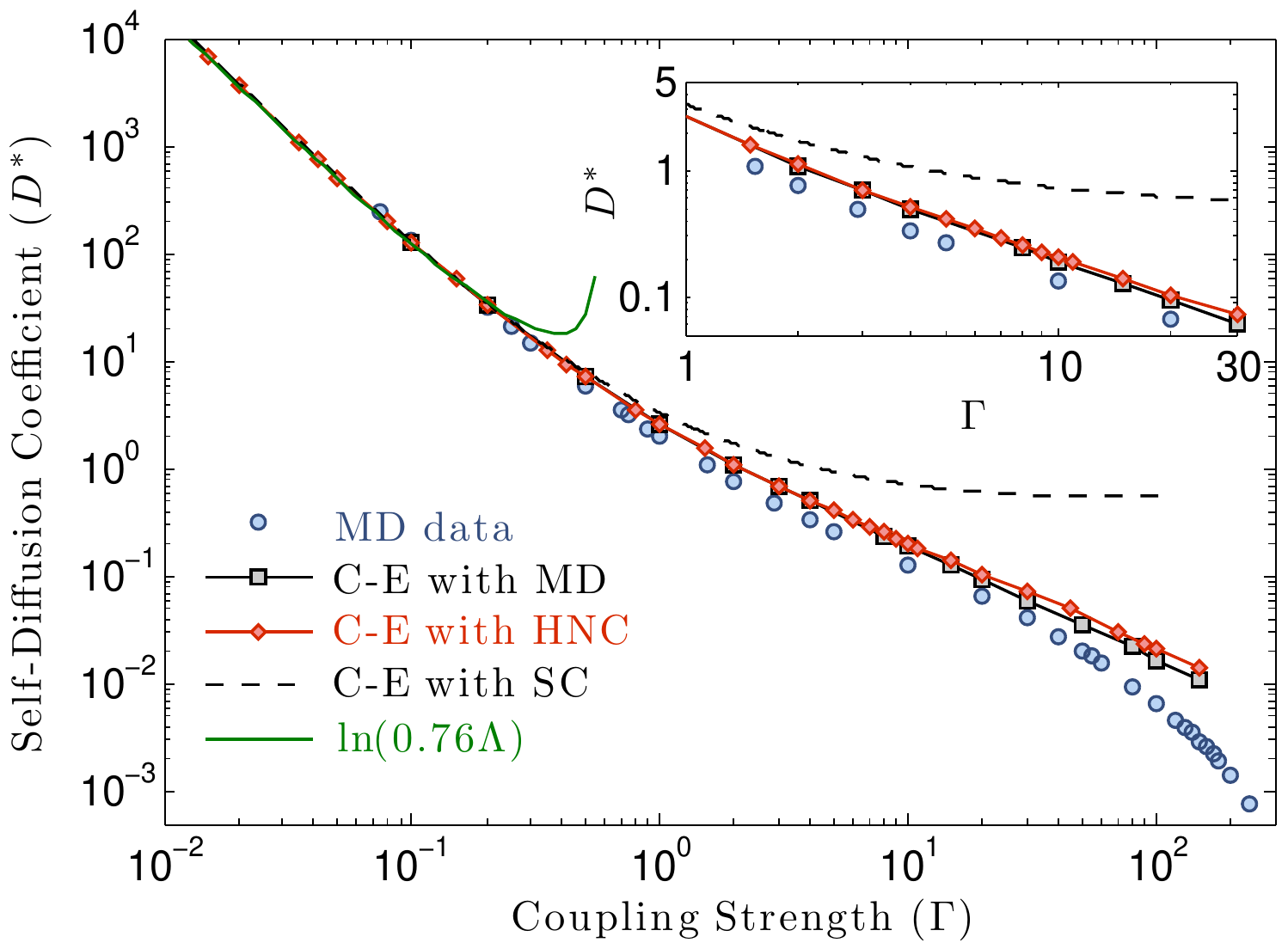}
\caption{Self diffusion coefficient for the OCP calculated using classical MD with Green-Kubo relations (blue circles), and using various effective potentials in the Chapman-Enskog collision integrals: from MD derived $g(r)$ data (black squares), HNC (red diamonds) and screened Coulomb (black dashed line).}
\label{fg:sd}
\end{figure}

\textit{Self diffusion in a OCP}: The Chapman-Enskog self diffusion coefficient to first order is \cite{chap:70}
\begin{equation}
[D_{ss^\prime}]_1 = \frac{3}{16} \frac{k_\B T}{n m_{ss^\prime} \Omega_{ss^\prime}^{(1,1)}}  .
\end{equation}
Accounting for a second order correction resulting from deviations from Maxwellian distributions provides
\begin{equation}
[D_{ss^\prime}]_2 = [D_{ss^\prime}]_1/(1 - \Delta) .
\end{equation}
where
\begin{equation}
\Delta = \frac{(2 \Omega_{ss^\prime}^{(1,2)} - 5 \Omega_{ss^\prime}^{(1,1)})^2/\Omega_{ss^\prime}^{(1,1)}}{55 \Omega_{ss^\prime}^{(1,1)} - 20 \Omega_{ss^\prime}^{(1,2)} + 4 \Omega_{ss^\prime}^{(1,3)} + 8 \Omega_{ss^\prime}^{(2,2)}} .
\end{equation}
For a OCP $s=s^\prime$. 

\begin{figure}
\includegraphics[width=8.0cm]{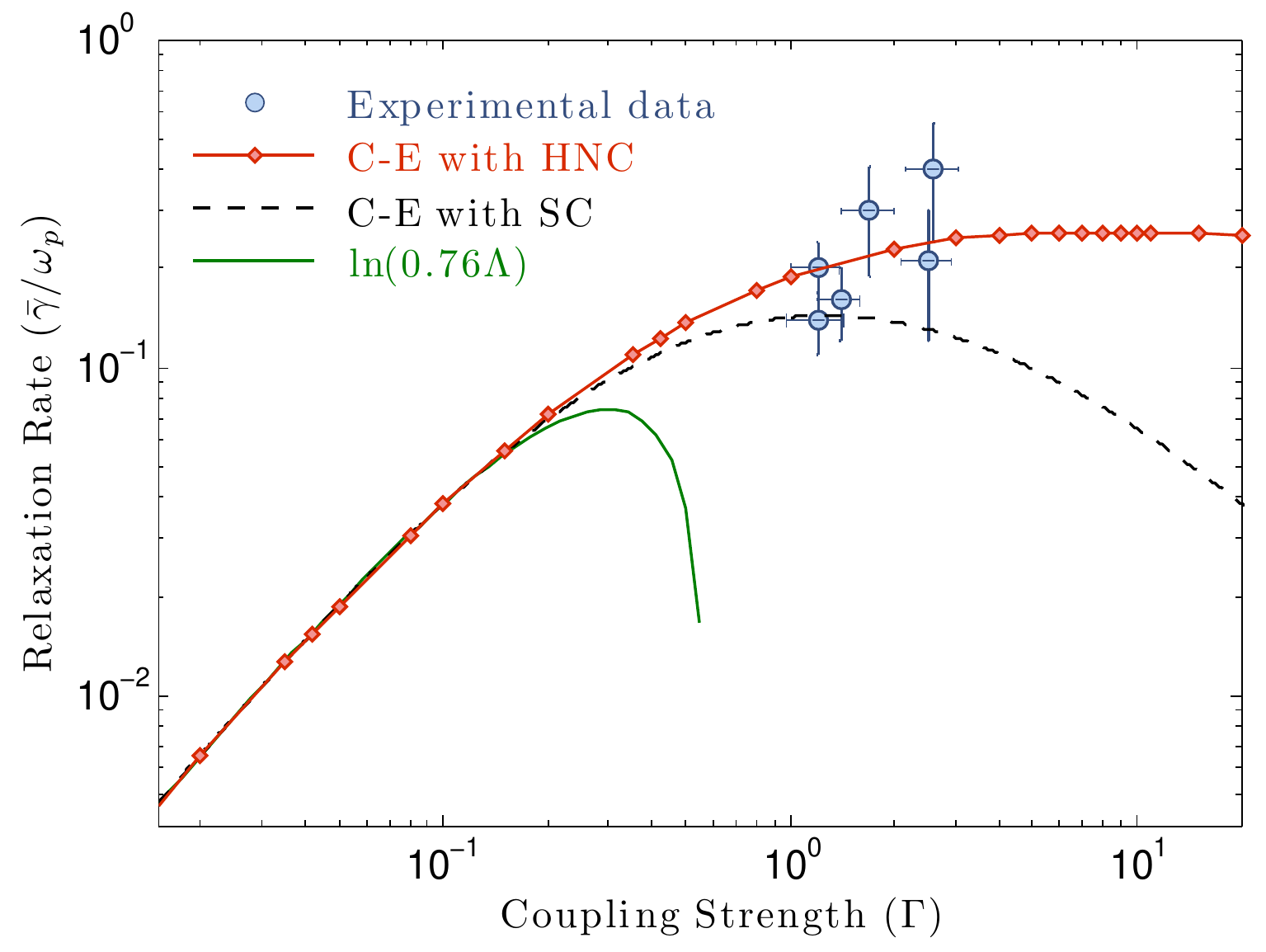}
\caption{Comparison of the velocity relaxation rate measured in \cite{bann:12} with theoretical predictions using HNC and screened Coulomb effective potentials. For a comparison of the experiment to other theories, see Fig.~5 of \cite{bann:12}.}
\label{fg:exp}
\end{figure}

Figure~\ref{fg:sd} shows a comparison between the self-diffusion coefficient obtained from using MD particle data in the Green-Kubo relations (using the method and code described in \cite{dali:12a}) and predictions obtained using the MD, HNC and screened Coulomb effective potentials in Eqs.~(\ref{eq:cints})--(\ref{eq:theta}). The figure shows excellent agreement for screened Coulomb in the weakly coupled limit, but this breaks down for $\Gamma \gtrsim 1$, as expected. Likewise, the HNC result is excellent in the weakly coupled regime, but this agreement also extends into the strongly coupled regime. This confirmation that the binary collision picture can be extended into the strongly coupled regime through the use of an effective potential that accounts for correlations is a primary result of this Letter. It provides a practical means for evaluating transport coefficients across coupling regimes. The results begin to diverge near the cross-over to the liquid regime~\cite{dali:12a}.  The curve obtained from using the MD extracted effective potential shows that only a small part of the disagreement between the direct MD data and effective potential theory comes from inadequacies of the HNC approximation.

\textit{Comparison with an experiment}: Recently Bannasch \emph{et al} \cite{bann:12} measured the velocity relaxation rate in a strongly coupled plasma. This experiment was conducted in an ultracold neutral plasma held in a magneto-optical trap and formed by photoionizing laser-cooled strontium atoms. Initially skewed velocity distributions for two spin states were formed using optical pumping, and the subsequent relaxation rate of the two distributions measured using laser induced fluorescence \cite{bann:12}. The ions in the system were in a strongly coupled regime, and the electrons formed a weakly coupled neutralizing background, providing a plasma in which the ion component is, to a good approximation, a classical OCP. 

Bannasch \emph{et al} extracted an average relaxation rate ($\bar{\gamma}$) from the time-resolved LIF measurements, and compared the results with predictions of previous theories that have the form, $\bar{\gamma}/\omega_p = 0.46 \Gamma^{3/2} \Xi$, where $\Xi$ is a generalized Coulomb logarithm. Figure~\ref{fg:exp} show a comparison of their data with theoretical predictions obtained using the HNC and screened Coulomb effective potentials to calculate $\Xi^{(1,1)}$. The figure shows that, within the measurement error, the effective potential theory agrees with the experiment when correlations are accounted for.

\begin{figure}
\includegraphics[width=8.0cm]{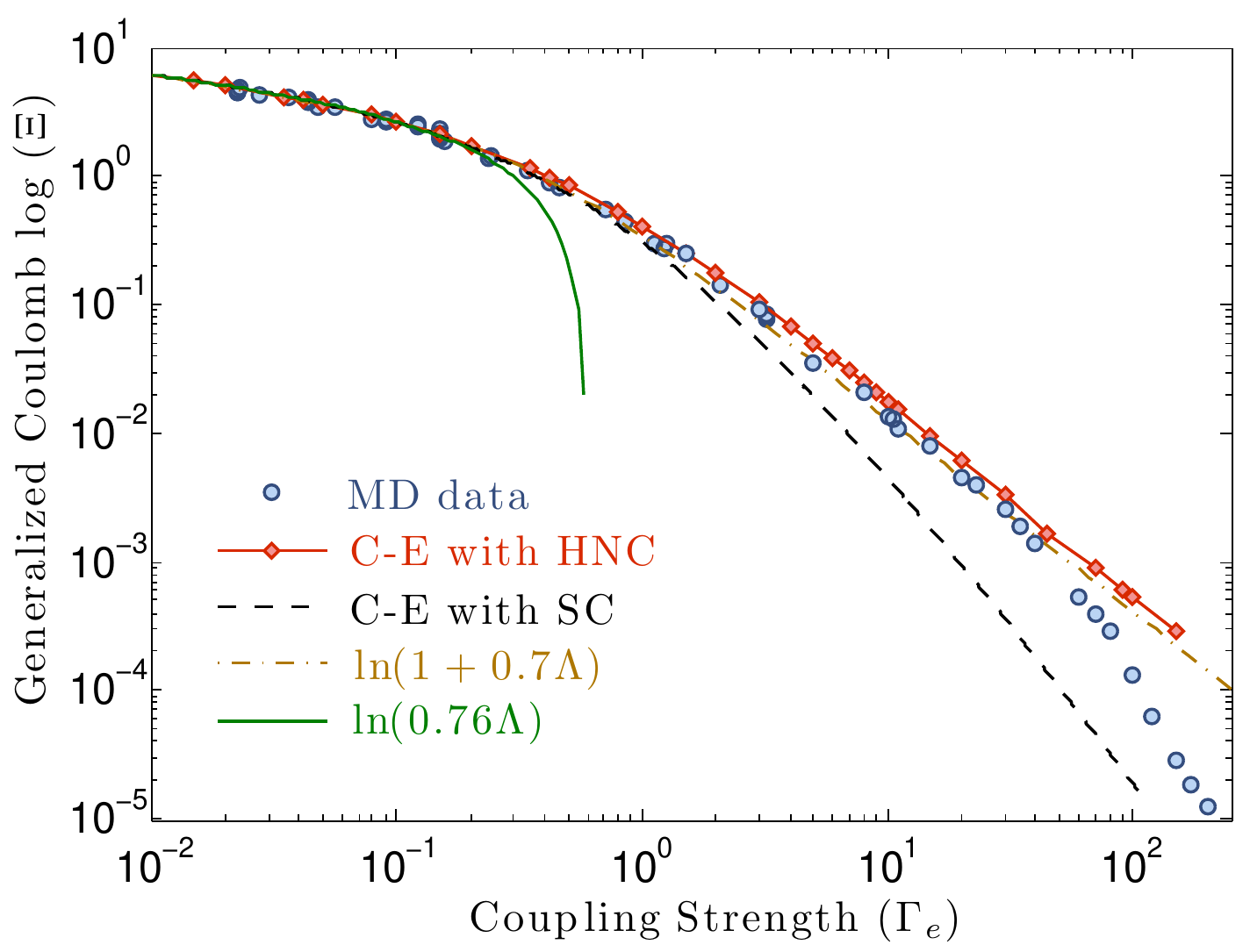}
\caption{A comparison of classical MD simulations and theoretical predictions for the generalized Coulomb logarithm in like-charge electron-ion thermal relaxation.}
\label{fg:tr}
\end{figure}

\textit{$e^{+} - i^+$ temperature equilibration}: Figure \ref{fg:tr} shows a comparison of the theoretical predictions and MD results for the like-charge electron-ion temperature relaxation rate. The generalized Coulomb logarithm is shown in the figure, which is inferred from $dT_e/dt = 2 Q^{e-i}/3n_e$ where $Q_{e-i} = -3 m_{ei} n_e \nu_{ei} (T_e - T_i)/m_i$ is the energy exchange density for Maxwellian distributions \cite{baal:12} and $\Xi$ enters through $\nu_{ei}$ from Eq.~(\ref{eq:nu}). Details of the MD simulations and analysis are explained in \cite{dimo:08}. The figure shows similar accuracy of the effective potential calculation for temperature relaxation across coupling regimes as was found for the self diffusion coefficient of an OCP.

Figures \ref{fg:sd}--\ref{fg:tr} show that the effective potential theory accurately predicts both experimental and \textit{ab initio} simulation data for a variety of different transport coefficients. This demonstrates both the flexibility of this approach (because it is compatible with the Chapman-Enskog formalism), and that the binary collision approximation can be extended into the strong coupling regime through the use of an effective potential that includes correlation effects. The approximation was shown to break down at very large $\Gamma$, where there is a known transition to liquid behavior and caging effects turn on~\cite{dali:12a}. Although the range of $\Gamma$ values for which this approximation is accurate is sufficient for many plasma physics applications, further refinements to the theory can also be envisioned. A potentially significant extension would be to account for a dynamic response function in both the closure that determines $g(r)$, and the relationship between the pair correlation function and the effective potential, Eq.~(\ref{eq:grphi}). This may provide an effective potential that accounts for relative particle velocities ($\vc{u}$), and associated wake effects.


The authors thank Dr.\ G.\ Bannasch for providing the experimental data points from \cite{bann:12} that were plotted in Fig.~\ref{fg:exp}. This research was supported under the auspices of the National Nuclear Security Administration of the U.S. Department of Energy at Los Alamos National Laboratory under Contract No. DE-AC52-06NA25396.


\bibliography{refs.bib}

\begin{thebibliography}{99}

\bibitem{chap:70} S.\ Chapman and T.\ Cowling, \emph{The Mathematical Theory of Non-Uniform Gases}, (Cambridge University, Cambridge, 1970).

\bibitem{brag:65} S.\ I.\ Braginskii, \textit{Reviews of Plasma Physics} vol.\ 1, ed. M.\ A.\ Leontovich (Consultants Bureau, New York, 1965).

\bibitem{land:36} L.\ Landau, Physik.\ Z.\ Sowjetunion {\bf 10}, 154 (1936). 

\bibitem{spit:53} L.\ Spitzer and R.\ Harm, Phys.\ Rev.\ {\bf 89}, 977 (1953); R.\ S.\ Cohen, L.\ Spitzer and P. McR.\ Routly, Phys.\ Rev.\ {\bf 80}, 230 (1950).

\bibitem{rose:57} M.\ N.\ Rosenbluth, W.\ M.\ MacDonald and D.\ L.\ Judd, Phys.\ Rev.\ {\bf 107}, 1 (1957).

\bibitem{lena:60} A.\ Lenard, Ann.\ Phys.\ (N.Y.) {\bf 3}, 390 (1960); R.\ Balescu, Phys.\ Fluids {\bf 3}, 52 (1960).

\bibitem{atze:04} S.\ Atzeni and J.\ Meyer-Ter-Vehn, {\it The Physics of Inertial Fusion} (Clarendon Press, Oxford, U.K., 2004).

\bibitem{lee:84} Y.\ T.\ Lee and R.\ M.\ More, Phys.\ Fluids {\bf 27}, 1273 (1984).

\bibitem{andr:10} G.\ B.\ Andresen, {\it et al}., Nature {\bf 468}, 673 (2010).

\bibitem{kill:07} T.\ C.\ Killian, Science {\bf 316}, 705 (2007).

\bibitem{merl:04} R.\ L.\ Merlino and J.\ A.\ Goree, Phys.\ Today {\bf 57}, 32 (2004).

\bibitem{pote:99} A.\ Y.\ Potekhin, Astron.\ Astrophys.\ {\bf 351}, 787 (1999).

\bibitem{much:84} D.\ Muchmore, Ap.\ J.\ {\bf 278}, 769 (1984).

\bibitem{iben:85} I.\ Iben and J.\ MacDonald, Ap.\ J.\ {\bf 296}, 540 (1985).

\bibitem{lore:09} W.\ Lorenzen, B.\ Holst and R.\ Redmer, Phys.\ Rev.\ Lett.\ {\bf 102}, 115701 (2009).

\bibitem{chab:06} G.\ Chabrier, D.\ Saumon and A.\ Y.\ Potekhin, J.\ Phys.\ A: Math.\ Gen.\ {\bf 39}, 4411 (2006).

\bibitem{dense} In the dense plasma examples, ions are typically in a classical regime, but electrons in a quantum regime. In this Letter, we focus only on the classical regime. 

\bibitem{baus:80} M.\ Baus and J.-P.\ Hansen, Phys.\ Reports {\bf 59}, 1 (1980).

\bibitem{desj:02} M.\ P.\ Desjarlais, J.\ D.\ Kress and L.\ A.\ Collins, Phys.\ Rev.\ E {\bf 66}, 025401 (2002).

\bibitem{donk:09} Z.\ Donk\'{o}, J.\ Phys.\ A: Math.\ Theor.\ {\bf 42}, 214029 (2009); and references cited therein

\bibitem{rose:95} Y.\ Rosenfeld, E.\ Nardi and Z.\ Zinamon, Phys.\ Rev.\ Lett.\ {\bf 75}, 2490 (1995).

\bibitem{rein:95} H.\ Reinholz, R.\ Redmer and S.\ Nagel, Phys.\ Rev.\ E {\bf 52}, 5368 (1995).

\bibitem{bene:12} L.\ X.\ Benedict, M.\ P.\ Surh, J.\ I.\ Castor, S.\ A.\ Khairallah, H.\ D.\ Whitley, D.\ F.\ Richards, J.\ N.\ Glosli, M.\ S.\ Murillo, C.\ R.\ Scullard, P.\ E.\ Grabowski, D.\ Michta and F.\ R.\ Graziani, Phys.\ Rev.\ E {\bf 86}, 046406 (2012). 

\bibitem{bann:12} G.\ Bannasch, J.\ Castro, P.\ McQuillen, T.\ Pohl, and T.\ C.\ Killian, Phys.\ Rev.\ Lett.\ {\bf 109}, 185008 (2012).

\bibitem{dimo:08} G.\ Dimonte and J.\ Daligault, Phys.\ Rev.\ Lett.\ {\bf 101}, 135001 (2008). 

\bibitem{dali:12a} J.\ Daligault, Phys.\ Rev.\ Lett.\ {\bf 108}, 225004 (2012); Phys.\ Rev.\ E {\bf 86}, 047401 (2012).

\bibitem{libo:59} R.\ L.\ Liboff, Phys.\ Fluids {\bf 2}, 40 (1959).

\bibitem{maso:67} E.\ A.\ Mason, R.\ J.\ Munn and F.\ J.\ Smith, Phys.\ Fluids {\bf 10}, 1827 (1967).

\bibitem{paqu:86} C.\ Paquette, C.\ Pelletier, G.\ Fontaine and G.\ Michaud, Astrophys.\ Journal Suppl.\ Series {\bf 61}, 177 (1986). 

\bibitem{baal:12} S.\ D.\ Baalrud, Phys.\ Plasmas {\bf 19}, 030701 (2012). 

\bibitem{geri:02} D.\ O.\ Gericke, M.\ S.\ Murillo and M.\ Schlanges, Phys.\ Rev.\ E {\bf 65}, 036418 (2002). 

\bibitem{ichi:92} S.\ Ichimaru, \emph{Statistical Plasma Physics: Vol.\ 1}, (Addison-Wesley, Tokyo, 1992).

\bibitem{dali:09} J.\ Daligault and G.\ Dimonte, Phys.\ Rev.\ E {\bf 79}, 056403 (2009). 

\bibitem{dali:11} J.\ Daligault, J.\ Stat.\ Phys.\ {\bf 143}, 1189 (2011).

\bibitem{boer:82} D.\ B.\ Boercker, F.\ J.\ Rogers and H.\ E.\ DeWitt, Phys.\ Rev.\ A {\bf 25}, 1623 (1982).

\bibitem{hans:06} J.-P.\ Hansen and I.\ R.\ McDonald, \emph{Theory of Simple Liquids, 3rd Edition} (Academic Press, Oxford, 2006).

\bibitem{hill:60} T.\ L.\ Hill, \emph{An Introdcution to Statistical Thermodynamics} (Addison-Wesley, Reading, 1960) p.\ 313.

\bibitem{epnote} The effective potential in Eq.~(\ref{eq:epdef}) has also been referred to as the potential of mean force~\cite{hill:60,hans:06}, and is sometimes denoted $\psi (1,2)$.  

\bibitem{aono:68} O.\ Aono, Phys.\ Fluids {\bf 11}, 341 (1968); and references cited therein 

\bibitem{lifs:81} E.\ M.\ Lifshitz and L.\ P.\ Pitaevskii, {\it Physical Kinetics} (Pergamon, Oxford, 1981) \S 46.

\bibitem{brow:05} L.\ S.\ Brown, D.\ L. Preston and R.\ L.\ Singleton Jr., Phys.\ Reports {\bf 410}, 237 (2005).


%





































\end{thebibliography}

\end{document}